# Effects of Competing Orders and Quantum Phase Fluctuations on the Low-Energy Excitations and Pseudogap Phenomena of Cuprate Superconductors

C.-T. Chen, A. D. Beyer, N.-C. Yeh,

*Department of Physics, California Institute of Technology, Pasadena, CA 91125, USA*

Received June 10 2007

**Abstract**

We investigate the low-energy quasiparticle excitation spectra of cuprate superconductors by incorporating both superconductivity (SC) and competing orders (CO) in the bare Green's function and quantum phase fluctuations in the proper self-energy. Our approach provides consistent explanations for various empirical observations, including the excess subgap quasiparticle density of states, "dichotomy" in the momentum-dependent quasiparticle coherence and the temperature-dependent gap evolution, and the presence (absence) of the low-energy pseudogap in hole- (electron-) type cuprates depending on the relative scale of the CO and SC energy gaps.

Cuprate superconductors differ fundamentally from conventional superconductors in that they are doped Mott insulators with strong electronic correlation that leads to possibilities of different competing orders (CO) in the ground state besides superconductivity (SC) [1-8]. The existence of competing orders and the proximity to quantum criticality [2, 3, 7, 8] gives rise to unconventional low-energy excitations of the cuprates, manifested as weakened superconducting phase stiffness [6], occurrence of excess subgap quasiparticle density of states (DOS) [9], spatial modulations in the low-temperature quasiparticle spectra that are unaccounted for by Bogoliubov quasiparticles alone [10-12], "dichotomy" in the momentum-dependent quasiparticle coherence [13] and temperature-dependent gap evolution, and the presence (absence) of the low-energy pseudogap (PG) [9, 15, 16] and Nernst effect [17] in the hole (electron)-type cuprates above the SC transition. Microscopically, the existence of CO is likely responsible for various non-universal phenomena among different cuprates [8, 9, 18, 19]. Macroscopically, the weakened superconducting phase stiffness and proximity to CO can give rise to strong fluctuations that lead to the extreme type-II nature and rich vortex dynamics [8, 20, 21].

To date there are two typical theoretical approaches to describing the quasiparticle excitation spectra of the cuprates. One approach takes the BCS-like Hamiltonian as the unperturbed mean-field state and a competing order, pinned by disorder, as the perturbation that gives rise to a weak scattering potential for the Bogoliubov quasiparticles [11, 22-24]. The other approach begins with the BCS-like Hamiltonian and includes superconducting phase fluctuations in the proper self-energy correction [25, 26]. However, no quantitative calculations have been made by incorporating both CO and quantum phase fluctuations in the SC state. The objective of this work is to consider the latter scenario and compute the corresponding low-energy excitation spectra with realistic physical parameters for comparison with experiments. We find that the low-energy excitations thus derived differ from typical Bogoliubov quasiparticles and can account for various puzzling phenomena aforementioned.

Our approach considers coexisting SC and CO with finite quantum phase fluctuations at zero temperature ($T = 0$) and employs realistic bandstructures for both electron- and hole-type cuprates [27, 28]. In addition, the temperature-dependent evolution of the quasiparticle low-energy excitation spectra is examined by comparing the results for coexisting SC and CO at $T = 0$ with those at $T > T_c$ in the mean-field limit. For the relevant competing orders, we focus on charge-density waves (CDW) and spin-density waves (SDW) in this work because of well documented empirical and theoretical evidences for their existence [5-8]. We assume that the density waves are static because dynamic density waves can be pinned by disorder. The momentum $\mathbf{k}$ remains a good quantum number as long as the mean free path is much longer than the superconducting coherence length, a condition generally satisfied in the cuprates. We further note that the spectroscopic characteristics associated with either CDW or disorder-pinned SDW as the CO are similar in the charge sector, although the wave-vector $\mathbf{Q}$ of CDW is twice of that of SDW with the direction of $\mathbf{Q}$ along the Cu-O bonding direction in the $CuO_2$ plane [22]. Regarding the SC pairing symmetry, in the case of hole-type cuprates we consider the empirically dominant scenario of coexisting $d_{x^2-y^2}$-wave SC with CDW or disorder-pinned SDW. For the electron-type cuprates we investigate two scenarios that are compatible with experimental findings: either $s$-wave SC and CDW or $d_{x^2-y^2}$-wave SC with disorder-pinned SDW. The former pairing symmetry is empirically justified from momentum-independent tunneling spectroscopy of the optimally doped infinite-layer cuprates [9], and the latter is verified with phase-sensitive experiments on certain one-layer cuprates [29]. It is also worth noting that both $s$-wave SC and CDW are symmetry representations of the SO(4) group [30], and their coexistence has been known in NbSe$_2$ [31]. In general, the key findings derived in this work can be extended to various pairing symmetries and other competing orders such as antiferromagnetism (AFM) [1] or $d$-density waves (DDW) [4]. As for the degree of quantum phase fluctuations, in the absence of known microscopic coupling mechanism between SC and CO, the magnitude of fluctuations is taken as a variable to be determined empirically.

The generalized mean field Hamiltonian for coexisting SC and CO is given by:

$$\mathcal{H}_{MF} = \mathcal{H}_{SC} + \mathcal{H}_{CO} = \sum_{\mathbf{k}} \Psi_{\mathbf{k},\sigma}^{\dagger} \mathcal{H}_0 \Psi_{\mathbf{k}}$$
$$= \sum_{\mathbf{k},\sigma} \xi_{\mathbf{k}} c_{\mathbf{k},\sigma}^{\dagger} c_{\mathbf{k},\sigma} - \sum_{\mathbf{k}} \Delta_{SC}(\mathbf{k}) \left( c_{\mathbf{k},\uparrow}^{\dagger} c_{-\mathbf{k},\downarrow}^{\dagger} + c_{-\mathbf{k},\downarrow} c_{\mathbf{k},\uparrow} \right)$$
$$- \sum_{\mathbf{k},\sigma} V_{CO} \left( c_{\mathbf{k},\sigma}^{\dagger} c_{\mathbf{k}+\mathbf{Q},\sigma} + c_{\mathbf{k}+\mathbf{Q},\sigma}^{\dagger} c_{\mathbf{k},\sigma} \right), \quad (1)$$

where $\xi_{\mathbf{k}}$ is the normal state energy of particles of momentum $\mathbf{k}$ relative to the Fermi energy, $c^{\dagger}$ and $c$ are the creation and annihilation operators, $\mathbf{Q}$ is the wave vector of the CO and its magnitude can be determined by the doping level for hole-type cuprates to satisfy the nesting condition [5], $\Delta_{SC}(\mathbf{k}) = \Delta_s$ for $s$-wave SC and $\Delta_{SC}(\mathbf{k}) = \Delta_d \cos(2\theta_k)$ for $d_{x^2-y^2}$-wave SC with $\theta_k$ being the angle between $\mathbf{k}$ and the anti-node of the pairing potential, $V_{CO}$ is the CO energy scale, $\sigma = \uparrow, \downarrow$ is the spin index, $\mathcal{H}_0$ is a $(4 \times 4)$ matrix, and $\Psi^{\dagger}$ represents a $(1 \times 4)$ matrix $\Psi^{\dagger} = \left( c_{\mathbf{k},\uparrow}^{\dagger} \; c_{-\mathbf{k},\downarrow} \; c_{\mathbf{k}+\mathbf{Q},\uparrow}^{\dagger} \; c_{-(\mathbf{k}+\mathbf{Q}),\downarrow} \right)$. In the case of CDW being the relevant CO, we have $V_{CO}(\mathbf{k}) = V_{CDW}$. For disorder-pinned SDW we express $V_{CO}(\mathbf{k}) = g^2 V_{SDW}$, where $V_{SDW}$ denotes the energy scale of SDW and $g$ is the coupling strength between SDW and disorder [22]. Here we have neglected the direct coupling of SDW to SC [32] in the Hamiltonian because the corresponding phase space contribution of the first-order SDW coupling to the DOS is too small in the doped cuprates, similar to the situation of negligible DDW coupling to the DOS in the doped cuprates, as elaborated in Refs. [33,34]. In principle, the CO energy $V_{CO}(\mathbf{k}) \equiv V_{CO} F(\mathbf{k})$ may contain a momentum-dependent form factor $F(\mathbf{k})$, as exemplified in Ref. [35] for a specific checkerboard CO pattern. In this work we consider a simple form factor $F(\mathbf{k})$ as a Gaussian distribution function centered at $\mathbf{k} = \mathbf{Q}$ so that $F(\mathbf{Q}) = 1$. This simple form factor serves the purpose of capturing the essence of the interplay between two energy scales $V_{CO}$ and $\Delta_{SC}$ without introducing excess adjustable parameters. Thus, the mean-field Hamiltonian in Eq. (1) can be exactly diagonalized so that the bare Green's function $G_0(\mathbf{k}, \omega)$ is given by $G_0^{-1} = \omega \mathbf{I} - \mathcal{H}_0$, where $\mathbf{I}$ denotes the $(4 \times 4)$ unit matrix.

Next, we introduce quantum phase fluctuations into the proper self-energy $\Sigma^*$. Specifically, from an effective low-energy theory [25, 26] the phase fluctuations for coexisting SC and CO can be evaluated through the velocity-velocity correlation function, such that

$$\Sigma^* = \sum_{\mathbf{q}} [m \mathbf{v}_g]_{\alpha} [m \mathbf{v}_g]_{\beta} \left\langle v_s^{\alpha}(\mathbf{q}) v_s^{\beta}(-\mathbf{q}) \right\rangle_{ring} G(\mathbf{k}-\mathbf{q}, \omega)$$
$$= \Sigma_l^*(\mathbf{k}, \omega) + \Sigma_t^*(\mathbf{k}, \omega)$$
$$= \sum_{\mathbf{q}} \left[ m \mathbf{v}_g(\mathbf{k}) \cdot \hat{\mathbf{q}} \right]^2 C_l(\mathbf{q}) G(\mathbf{k}-\mathbf{q}, \omega)$$
$$+ \sum_{\mathbf{q}} \left[ m \mathbf{v}_g(\mathbf{k}) \times \hat{\mathbf{q}} \right]^2 C_t(\mathbf{q}) G(\mathbf{k}-\mathbf{q}, \omega). \quad (2)$$

In Eq. (2), realistic bandstructures $\xi_{\mathbf{k}}$ of the cuprates are incorporated explicitly by expressing the group velocity as $m_e \mathbf{v}_g = (\partial \xi_{\mathbf{k}} / \partial \mathbf{k}) / \hbar$ for $|\mathbf{k}| \sim k_F$. At $T = 0$ and in zero magnetic field, the longitudinal phase fluctuations dominate so that $\Sigma^* = \Sigma_l^* + \Sigma_t^* \approx \Sigma_l^*$. Summing over an infinite series of ring diagrams [25], we arrive at the expression for the coefficient $C_l(\mathbf{q})$ in $\Sigma_l^*$ [26]:

$$\left| C_l(\mathbf{q}) \right| = \frac{\hbar^2 q^2}{4 m_e^2} \left\langle \theta(\mathbf{q}) \theta(-\mathbf{q}) \right\rangle \equiv \frac{4 \hbar^2 q^2}{m_e^2} \eta, \quad (3)$$

where $\theta$ is the SC phase renormalized by the presence of CO, $\langle \ldots \rangle$ denotes the correlation function, $m_e$ the free electron mass, and $\eta$ is a dimensionless parameter indicative of the magnitude of phase fluctuations. In general, $\eta$ consists of contributions from both SC and CO, and its mean value upon averaging over the Fermi surface yields $(\hbar \omega_p / n_{s,2D} m_e) \equiv \eta$, where $\omega_p$ denotes the plasma frequency and $n_{s,2D}$ represents the effective two-dimensional superfluid density [26]. Finally, to build into our model realistic finite lifetime broadenings for quasiparticles due to impurity scattering and quantum phase fluctuations, we relax the condition for the imaginary part of the inverse bare Green's function $G_0^{-1}$ from $0^+$ to a finite but small quantity $\delta(\mathbf{k})$. Specifically, we express $\delta(\mathbf{k}) = \delta D(\mathbf{k})$ with $\delta \ll \Delta_{eff}$. If we further assume that the quasiparticle lifetime broadening is primarily associated with quantum phase fluctuations, the form factor $D(\mathbf{k})$ for the quasiparticle lifetime broadening becomes related to a mean-field effective gap $\Delta_{eff}(\mathbf{k}) = \left[ \Delta_{SC}^2(\mathbf{k}) + V_{CO}^2(\mathbf{k}) \right]^{1/2} \equiv \Delta_{eff} D(\mathbf{k})$, where $\Delta_{eff} = \left( \Delta_{SC}^2 + V_{CO}^2 \right)^{1/2}$ is the maximum effective gap. Thus, we obtain the full Green's function $G(\mathbf{k}, \tilde{\omega})$ self-consistently through the Dyson's equation (with $\xi_{\mathbf{k}} \approx -\xi_{\mathbf{k}+\mathbf{Q}}$ for the particle-hole excitations in the CO density-wave channel):

$$G^{-1}(\mathbf{k}, \tilde{\omega}) = G_0^{-1}(\mathbf{k}, \tilde{\omega}) - \sum_{\mathbf{q}} \left[ m \mathbf{v}_g(\mathbf{k}) \cdot \hat{\mathbf{q}} \right]^2 C_l(\mathbf{q}) G(\mathbf{k}-\mathbf{q}, \tilde{\omega})$$
$$= \begin{pmatrix} \tilde{\omega} - \tilde{\xi}_{\mathbf{k}} & \tilde{\Delta}_{\mathbf{k}} & \tilde{V}_{\mathbf{k}} & 0 \\ \tilde{\Delta}_{\mathbf{k}} & \tilde{\omega} + \tilde{\xi}_{\mathbf{k}} & 0 & -\tilde{V}_{\mathbf{k}} \\ \tilde{V}_{\mathbf{k}} & 0 & \tilde{\omega} + \tilde{\xi}_{\mathbf{k}} & \tilde{\Delta}_{\mathbf{k}+\mathbf{Q}} \\ 0 & -\tilde{V}_{\mathbf{k}} & \tilde{\Delta}_{\mathbf{k}+\mathbf{Q}} & \tilde{\omega} - \tilde{\xi}_{\mathbf{k}} \end{pmatrix}. \quad (4)$$

Here $\tilde{\omega}(\mathbf{k}, \omega)$ denotes the energy renormalized by the phase fluctuations. Equation (4) is solved self-consistently by first choosing an energy $\omega$, going over the $\mathbf{k}$-values in the Brillouin zone by summing over a finite phase space in $\mathbf{q}$ near each $\mathbf{k}$, and then finding the corresponding fluctuation-renormalized $\tilde{\xi}_{\mathbf{k}}$, $\tilde{\omega}$ and $\tilde{\Delta}_{SC}$ until the solution to the full Green's function $G(\mathbf{k}, \tilde{\omega})$ converges iteratively. The spectral density function is given by $A(\mathbf{k}, \omega) \equiv -\mathrm{Im}\left[G(\mathbf{k}, \tilde{\omega}(\mathbf{k}, \omega))\right]/\pi$ and the DOS by $\mathcal{N}(\omega) = \sum_{\mathbf{k}} A(\mathbf{k}, \omega)$. The conservation of spectral weight is also confirmed throughout the calculations.

For a given cuprate bandstructure with a known doping level and pairing symmetry, both $A(\mathbf{k}, \omega)$ and $\mathcal{N}(\omega)$ calculated from our approach are determined by five parameters: ($\Delta_{SC}$, $V_{CO}$, $\mathbf{Q}$, $\eta$, $\delta$). For instance, the energy integrated $A(\mathbf{k}, \omega)$ may be compared with the angle-resolved photoemission spectroscopy (ARPES) in the first Brillouin zone to determine the CO wave-vector $\mathbf{Q}$. Similarly, the parameters ($\Delta_{SC}$, $V_{CO}$, $\mathbf{Q}$, $\eta$, $\delta$) obtained from best theoretical fitting to empirical $\mathcal{N}(\omega)$ data are in principle uniquely determined within experimental errors because of strong physical constraints on all parameters: the values of $\Delta_{SC}$ and $V_{CO}$ are simultaneously constrained by the spectral peak positions; the relative height of the spectral peaks depends on $\mathbf{Q}$, and $\mathbf{Q}$ thus determined can be cross-checked with ARPES data; the magnitude of $\eta$ is directly correlated with the DOS near $\omega = 0$; and the lifetime broadening $\delta$ is associated with the linewidth of the spectral peaks and the spectral slope of the DOS near $\omega = 0$. Based on the methods outlined above and comparison with experimental data, we summarize below four key findings.

First, the spectral density function $A(\mathbf{k}, \omega)$ broadens with increasing magnitude of quantum phase fluctuations $\eta$, and both the quasiparticle coherence, as manifested by the inverse linewidth ($\Gamma^{-1}$) of $A(\mathbf{k}, \omega)$, and the renormalized effective gap $\tilde{\Delta}_{eff}(\mathbf{k})$, exhibit "dichotomy" in the momentum space, showing different evolution in the Cu-O bonding direction $(0, \pi)/(\pi, 0)$ from that in the $(\pi, \pi)$ nodal direction, as confirmed by recent ARPES results [14]. Specifically, the fluctuation-renormalized $A(\mathbf{k}, \omega)$-vs.-$\omega$ (solid lines) is compared with the mean-field $A(\mathbf{k}, \omega)$ (dashed lines, $\eta = 0$) in Figs. 1(a) and (b) for an electron-type cuprate with coexisting $s$-wave SC and CDW, where $\mathbf{k}$ is respectively along $(0, \pi)/(\pi, 0)$ and $(\pi, \pi)$, and in Figs. 1(d) and (e) for a hole-type cuprate with coexisting $d_{x^2-y^2}$-wave SC and disorder-pinned SDW. We find that the most significant fluctuation-induced broadening in $A(\mathbf{k}, \omega)$ occurs at the Fermi level, which corresponds to $\xi_{\mathbf{k}} = 0$ or equivalently $\omega \sim \tilde{\Delta}_{eff}(\mathbf{k})$. Moreover, $\tilde{\Delta}_{eff}(\mathbf{k})$ is momentum dependent regardless of the pairing symmetry, although the $\mathbf{k}$-dependence is more pronounced for $d_{x^2-y^2}$-wave SC, as illustrated in Fig. 1(c) for the first Brillouin zone of the electron-type cuprate with $s$-wave SC/CDW and in Fig. 1(f) for that of the hole-type cuprate with $d_{x^2-y^2}$-wave SC/SDW.

In addition to the $\mathbf{k}$-dependent effective gap, dichotomy in the quasiparticle coherence can be manifested by comparing the linewidth of $A(\mathbf{k}, \omega \sim \tilde{\Delta}_{eff})$ for $\mathbf{k}$ along $(\pi, \pi)$ with that for $\mathbf{k}$ along $(0, \pi)/(\pi, 0)$, as exemplified in Figs. 2(a) and 2(b) for $s$-wave SC/CDW and $d_{x^2-y^2}$-wave SC/SDW, respectively. The degree of dichotomy in the quasiparticle coherence decreases with increasing quantum fluctuations, as shown in the inverse linewidth $\Gamma^{-1}$-vs.-$\eta$ plots in Figs. 2(c) and 2(d). In particular, we note that quasiparticles of $d_{x^2-y^2}$-wave SC/SDW exhibit better coherence along $(\pi, \pi)$ than along $(0, \pi)/(\pi, 0)$ for small $\eta$, which is consistent with ARPES data [13].

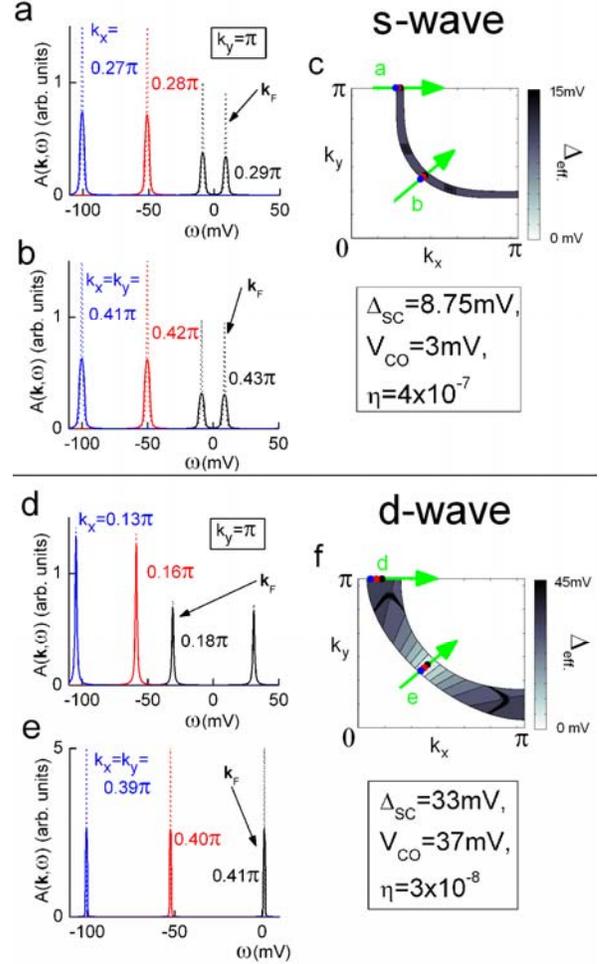

Fig. 1. (Color online) Dichotomy in the spectral density function and excitation gap due to coexisting SC/CO and quantum phase fluctuations: The fluctuation-renormalized $A(\mathbf{k}, \omega)$ (solid lines) and the mean-field $A(\mathbf{k}, \omega)$ (dashed lines, with $\eta = 0$) are illustrated for $\mathbf{k}$ along (a) $(\pi, 0)/(0, \pi)$ and (b) $(\pi, \pi)$ of an electron-type cuprate with coexisting $s$-wave SC and CDW, as illustrated by the arrows in (c) for the first Brillouin zone (BZ) of the $s$-wave SC/CDW, and for $\mathbf{k}$ along (d) $(\pi, 0)/(0, \pi)$ and (e) $(\pi, \pi)$ of a hole-type with coexisting $d_{x^2-y^2}$-wave SC and disorder-pinned SDW, as illustrated by the arrows in (f) for the first BZ of the $d_{x^2-y^2}$-wave SC/SDW. Here we have taken the CO wave-vector $\mathbf{Q}$ to be along $(\pi, 0)/(0, \pi)$ direction and the magnitude of $\mathbf{Q}$ to be nested with the Fermi surface. The renormalized effective gap $\tilde{\Delta}_{eff}(\mathbf{k})$ in the first BZ for states with $|\xi_{\mathbf{k}}| \leq 75$ meV of the $s$-wave SC/CDW is shown in (c), and that for states with $|\xi_{\mathbf{k}}| \leq 220$ meV of the $d_{x^2-y^2}$-wave SC/SDW is shown in (f), revealing $\mathbf{k}$-dependence regardless of the pairing symmetry, although the degree of dichotomy in the $d_{x^2-y^2}$-wave SC/SDW is apparently more significant.

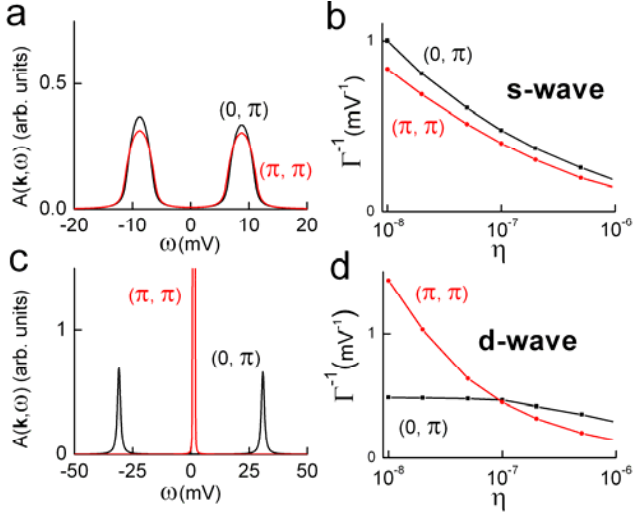

Fig. 2. (Color online) Contrasts in the fluctuation renormalized $A(\mathbf{k},\omega)$ at the Fermi level for $\mathbf{k}$ along $(\pi,0)/(0,\pi)$ [darker (black) lines] and along $(\pi,\pi)$ [lighter (red) lines] are shown in **(a)** for $s$-wave SC/CDW with $\eta = 4 \times 10^{-7}$, and in the **(b)** for $d_{x^2-y^2}$-wave SC/SDW with $\eta = 3 \times 10^{-8}$. Given that a broader linewidth $\Gamma$ (or equivalently, smaller $\Gamma^{-1}$) represents reduced quasiparticle coherence, we find that for $s$-wave SC/CDW the quasiparticles along $(\pi,0)/(0,\pi)$ are always more coherent than those along $(\pi,\pi)$ for all fluctuations considered, as shown by the $\Gamma^{-1}$ vs. $\eta$ plot in **(c)**. In contrast, for $d_{x^2-y^2}$-wave SC/SDW the quasiparticles along $(\pi,\pi)$ are more coherent than those along $(\pi,0)/(0,\pi)$ if the quantum fluctuations are sufficiently small, as manifested by the $\Gamma^{-1}$ vs. $\eta$ plot in **(d)**. The latter finding is consistent with the empirical observation of more coherent nodal quasiparticles in hole-type cuprate superconductors [13, 14].

Second, we find that the CO wave-vector $\mathbf{Q}$ need not be commensurate with $(\pi/a)$ to have effect on the low-energy excitations, although maximum effect occurs when $|\mathbf{k}|, |\mathbf{k}\pm\mathbf{Q}| \sim k_F$ [33]. Further comparison of our calculations with experimental APRES and DOS data suggests that $\mathbf{Q}$ in hole-type cuprates is best described in terms of either a doping-dependent and incommensurate CDW or disorder-pinned SDW. In contrast, best theoretical fitting to empirical results in electron-type cuprates implies that $\mathbf{Q}$ is commensurate and doping independent [33]. Thus, theoretical analysis of the $\mathbf{Q}$-dependence in the quasiparticle excitation spectra can provide information about whether a CO phase is relevant to the observed APRES and DOS [33]. For instance, we find that DDW with $\mathbf{Q}=(\pi,\pi)$ only makes significant contributions to the DOS in the insulating limit [33, 34]. Further details related to the effect of the CO wave-vector can be found in Ref. [33].

Third, finite quantum phase fluctuations can induce excess subgap DOS if $V_{CO} < \Delta_{SC}$, as illustrated in Fig. 3(a) for the comparison of calculated results with the momentum-independent quasiparticle tunneling spectra of the infinite-layer electron-type cuprate $Sr_{0.9}La_{0.1}CuO_2$ [9]. The reasonable agreement of our calculations with experimental data is obtained by assuming coexisting $s$-wave SC and CDW, with $\Delta_{SC} = 8.75$ meV, $V_{CO} = 8.75$ meV, $Q = 2\pi/3$ and $\eta = 2 \times 10^{-6}$.

On the other hand, for $d_{x^2-y^2}$-SC/SDW with $V_{CO} = 37$ meV $> \Delta_{SC} = 33$ meV, we find that the incorporation of CO can account for the two sets of peak features in the c-axis tunneling data of $Bi_2Sr_2CaCu_2O_x$ (Bi-2212) at $T \ll T_c$, as exemplified in Fig. 3(b). We further note that significant spectral variations can be induced by varying the quantum fluctuations even for fixed $\Delta_{SC}$ and $V_{CO}$, as shown in the inset of Fig. 3(b). For completeness, we also show in Fig. 3(c) the calculated DOS for coexisting $d_{x^2-y^2}$-wave SC and disorder-pinned SDW together with the empirical DOS taken on a 24° tilt (001) grain-boundary junction of a one-layer electron-type cuprate $Pr_{1.85}Ce_{0.15}CuO_{4-y}$ [16]. By using the parameters $\Delta_{SC} = 4.75$ meV and $V_{CO} = 4.75$ meV, we find that the calculated DOS is consistent with the spectrum in Fig. 3(a) that reveals only one set of peaks at $\Delta_{eff}$ under the condition $V_{CO} < \Delta_{SC}$, regardless of the SC pairing symmetry.

Fourth, we find that whether the low-energy PG occurs is primarily determined by the ratio of $\Delta_{SC}$ to $V_{CO}$: For arbitrary values of $\Delta_{SC}$ and $V_{CO}$, the poles associated with $\mathcal{H}_{MF}$ in Eq. (1) generally give rise to two sets of peaks at $\omega = \pm\tilde{\Delta}_{eff}$ and $\omega = \pm\tilde{\Delta}_{SC}$ in the quasiparticle DOS. If $\Delta_{SC} >> V_{CO}$ and $\eta > 0$, we have $\tilde{\Delta}_{eff} \approx \tilde{\Delta}_{SC}$ so that only one set of peaks can be resolved in the quasiparticle spectra. Hence, the magnitude of $\tilde{\Delta}_{eff}$ in the quasiparticle spectra decreases with increasing $T$ and vanishes at $T_c$, which is analogous to the behavior of conventional superconductors and is also consistent with all empirical findings to date in the electron-type cuprate superconductors [9, 16, 17], as exemplified in the main panel of Fig. 3(a). On the other hand, in the case of $V_{CO} > \Delta_{SC}$ two distinct sets of peaks can be resolved at $T << T_c$ even under moderate quantum phase fluctuations, as exemplified in the main panel of Fig. 3(b). With increasing $T$ or $\eta$, the DOS peaks at $\omega = \pm\tilde{\Delta}_{SC}$ steadily diminish while the peaks at $\omega = \pm\tilde{\Delta}_{eff} \approx \pm V_{CO}$ are broadened by fluctuations but the peak positions remain nearly invariant above $T_c$ as long as $k_B T << V_{CO}$. To better illustrate this point, the evolution of the quasiparticle tunneling spectra with varying ($\Delta_{SC}/V_{CO}$) and under a fixed $\eta$ at $T = 0$ is exemplified in Fig. 4(a) for an $s$-wave SC/CDW and in Fig. 4(b) for a $d_{x^2-y^2}$-wave SC/SDW. In addition, we show in the inset of Fig. 4(b) the comparison of calculated mean-field quasiparticle tunneling spectra at $T = 0$ and $T = 100$ K $> \sim T_c$ in the case of $V_{CO} = 50$ meV $> \Delta_{SC} = 33$ meV, which reveals thermally smeared PG features at $|\omega| \sim V_{CO}$ for $T > \sim T_c$ [39]. This finding is similar to the empirically observed low-energy PG phenomena in underdoped Bi-2212 cuprates above $T_c$ [15], and is also in sharp contrast to the "two-gap" features at $|\omega| = \tilde{\Delta}_{eff}$ and $\tilde{\Delta}_{eff}$ for $T = 0$. These results therefore suggest that the experimental observation of non-universal low-energy PG phenomena in the cuprates may be reconciled by the coexistence of CO and SC with different relative strengths. Thus, the absence of low-energy PG in electron-type cuprates can be attributed to $V_{CO} < \Delta_{SC}$, whereas the presence of low-energy PG in under- and optimally doped hole-type cuprates is due to $V_{CO} > \Delta_{SC}$.

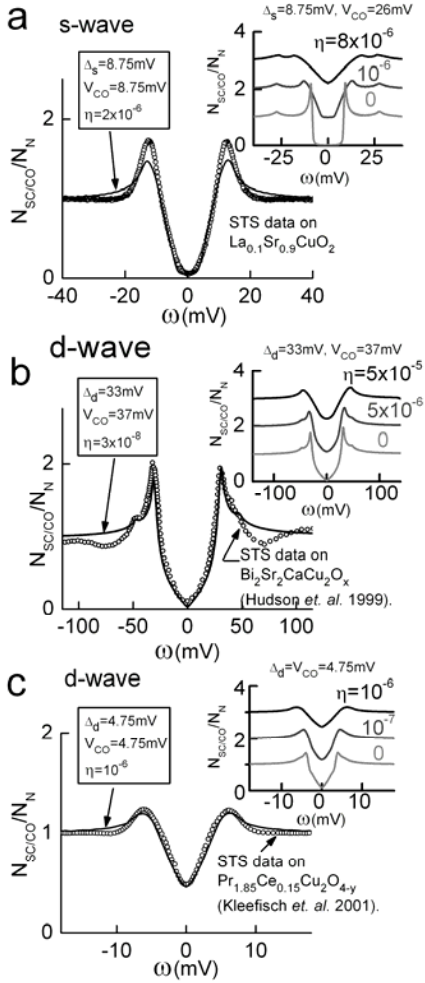

Fig. 3. (a) Representative momentum-independent quasiparticle tunneling spectrum (taken at 4.2 K) of the infinite-layer electron-type cuprate $Sr_{0.9}La_{0.1}CuO_2$ in Ref. [9] (open circles) is compared with the calculated DOS by assuming coexisting s-wave SC/CDW. The excess subgap DOS observed empirically can be accounted for by incorporating the condition of $V_{CO} = \Delta_{SC} = 8.75$ meV and finite quantum fluctuations. The inset shows the spectral evolution with increasing quantum fluctuations for fixed values of $\Delta_{SC}$ and $V_{CO}$, with $\eta$ values from bottom curve up being $(0, 1, 8) \times 10^{-6}$, and the curves have been uniformly up-shifted for clarity. (b) Quasiparticle c-axis tunneling spectrum of the hole-doped $d_{x^2-y^2}$-wave superconductor Bi-2212 in Ref. [36] (open circles) is compared with calculated DOS. The PG feature can be reproduced by the condition $V_{CO} > \Delta_{SC}$ and $\eta > 0$. The inset shows how the spectrum evolves with increasing $\eta$ for fixed values of $\Delta_{SC}$ and $V_{CO}$, which resembles certain local spatial variations in the quasiparticle tunneling spectra of Bi-2212 [37]. The $\eta$ values for the curves from bottom up are $(0, 5, 50) \times 10^{-6}$. The pronounced "dip" features unaccounted for by our calculations may be attributed to higher-energy bosonic excitations such as spin fluctuations [38]. (c) Quasiparticle tunneling spectrum taken on a grain-boundary junction of a one-layer optimally doped electron-type cuprate $Pr_{1.85}Ce_{0.15}CuO_{4-y}$ [16] is compared with calculated DOS for coexisting $d_{x^2-y^2}$-wave SC/SDW. The inset shows the effect of increasing $\eta$ with $\eta = (0, 1, 10) \times 10^{-7}$ from bottom up.

We remark that our finding of CO being responsible for the low-energy PG phenomena differs fundamentally from the conjecture of PG being a strongly phase fluctuating SC state above $T_c$ [25]. In the latter scenario, PG evolves into SC upon lowering the temperature so that there is only one SC ground state. This scenario of phase fluctuations solely arising from SC cannot account for the empirical two-gap features at $T << T_c$, nor can it explain the quasiparticle spectral dichotomy [13, 14] and the existence of CO below $T_c$ [7, 11].

A feasible mechanism leading to varying $(\Delta_{SC}/V_{CO})$ ratio in different cuprates is the coupling of charge to the longitudinal optical (LO) phonon mode in the $CuO_2$ plane: Given that strong ligand-hole hybridization occurs in the hole-type cuprates, the charge transfer of holes along the Cu-O bond can be enhanced by the LO phonon mode in the underdoped hole-type cuprates because the slower hole hopping rate in underdoped cuprates can better couple to the LO phonons [40]. In contrast, for electron-type cuprates, the charge transfer gap along the Cu-O bond remains large upon electron doping so that the LO phonons cannot effectively assist charge transfer. Hence, if CDW (or SDW-induced CDW) is the relevant CO, LO phonons associated with the Cu-O bonds can enhance the strength of CO in the underdoped hole-type cuprates [40], leading to an increasing isotope effect and stronger low-energy PG phenomena with decreasing hole doping [33]. Finally, we suggest that quantum phase fluctuations between CO and SC may contribute to the observed Nernst effect for magnetic field $H > H_{c2}$ in underdoped hole-type cuprates where $V_{CO} > \Delta_{SC}$, because at $T << T_c$ quantum phase fluctuations may induce SC from CO, thereby supporting vortices even at $H > H_{c2}$. Similarly, for temperature above $T_c$ and below the low-energy PG temperature, fluctuating SC may survive in an otherwise predominantly CO phase via the "giant proximity effect" [41] so that vortices can exist above $T_c$, leading to observation of the Nernst effect. This notion of coexisting "puddles" of SC and AFM (with AFM being the relevant CO) has been investigated numerically in Ref. [42].

In summary, we have investigated the physical origin of unconventional quasiparticle excitations and the low-energy pseudogap (PG) phenomena in cuprate superconductors by considering the effect of coexisting competing orders (CO) and superconductivity (SC). With explicit incorporation of both SC and CO in the bare Green's function and of quantum phase fluctuations in the proper self-energy, we can consistently account for various spectral characteristics associated with the ARPES and DOS data in both hole- and electron-type cuprates. In particular, we attribute the presence (absence) of the low-energy PG phenomena in hole-type (electron-type) cuprate superconductors to the competition of two energy scales, $V_{CO}$ for CO and $\Delta_{SC}$ for SC, with the condition $V_{CO} > \Delta_{SC}$ being responsible for the occurrence of low-energy PG above $T_c$.

### Acknowledgements

This work was supported by NSF Grant DMR-0405088.

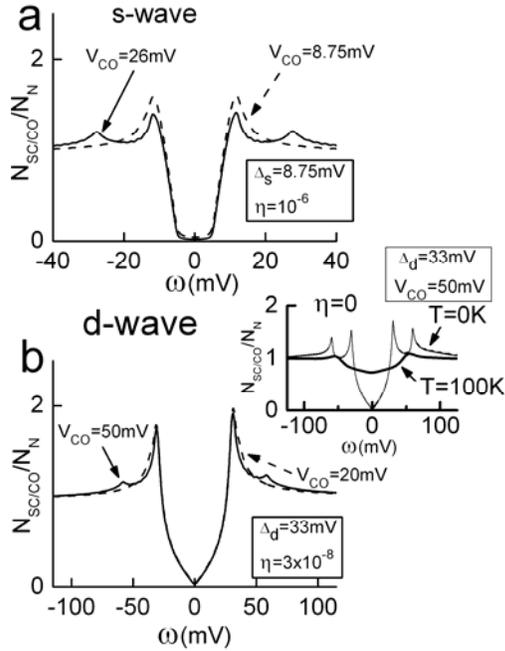

Fig. 4. A scenario for the occurrence of the low-energy PG as the consequence of coexisting CO and SC with $V_{CO} > \Delta_{SC}$: **(a)** Evolution of the quasiparticle tunneling spectra at $T = 0$ for *s*-wave SC/CDW from absence of PG to appearance of PG with increasing $(V_{CO}/\Delta_{SC})$ and under a fixed $\eta$. **(b)** Evolution of the *c*-axis quasiparticle tunneling spectra at $T = 0$ for a $d_{x^2-y^2}$-wave SC/SDW from absence to appearance of PG with increasing $(V_{CO}/\Delta_{SC})$ and under a fixed $\eta$. The inset shows the comparison of the mean-field normalized tunneling conductance spectrum with $V_{CO} = 50$ meV at $T = 0$ (thick line) and that at $T = 100$ K $> \sim T_c \sim 92$ K (thin line). Here we have used the condition $(dI/dV) \propto |\int \mathcal{N}(E) [df(E)/dE]_{E-eV} dE|$, and the Fermi function is given by $f(E) = [1 + \exp(-E/k_BT)]^{-1}$. Apparently only the thermally smeared PG features at $|\omega| \sim V_{CO}$ persists above $T_c$.